\documentclass[12pt]{iopart}
\usepackage{graphicx,color}
\begin{document}

\comment[Comment on `A scattering quantum circuit for measuring Bell's time inequality']{Comment on `A scattering quantum circuit for measuring Bell's time inequality: a nuclear magnetic resonance demonstration using maximally mixed states'}

% repeat the \author .. \affiliation  etc. as needed
% \email, \thanks, \homepage, \altaffiliation all apply to the current
% author. Explanatory text should go in the []'s, actual e-mail
% address or url should go in the {}'s for \email and \homepage.
% Please use the appropriate macro foreach each type of information

% \affiliation command applies to all authors since the last
% \affiliation command. The \affiliation command should follow the
% other information
% \affiliation can be followed by \email, \homepage, \thanks as well.
\author{G~C~Knee$^1$, E~M~Gauger$^{2,1}$, G~A~D~Briggs$^1$, S~C~Benjamin$^{1,2}$}
%\email[]{Your e-mail address}
%\homepage[]{Your web page}
%\thanks{}
%\altaffiliation{}
\ead{george.knee@materials.ox.ac.uk}
\address{$^1$ Department of Materials, University of Oxford, Parks Road, Oxford, OX1 3PH}
\address{ $^2$ Centre for Quantum Technologies, National University of Singapore, 3 Science Drive 2, Singapore 117543}

%Collaboration name if desired (requires use of superscriptaddress
%option in \documentclass). \noaffiliation is required (may also be
%used with the \author command).
%\collaboration can be followed by \email, \homepage, \thanks as well.
%\collaboration{}
%\noaffiliation

\date{\today}

\begin{abstract}
A recent paper by Souza, Oliveira and Sarthour (SOS) reports the experimental violation of a Leggett-Garg (LG) inequality (sometimes referred to as a temporal Bell inequality). The inequality tests for quantum mechanical superposition: if the inequality is violated, the dynamics cannot be explained by a large class of classical theories under the heading of macrorealism. Experimental tests of the LG inequality are beset by the difficulty of carrying out the necessary so-called `non-invasive' measurements (which for the macrorealist will extract information from a system of interest without disturbing it). SOS argue that they nevertheless achieve this difficult goal by putting the system in a maximally mixed state. The system then allegedly undergoes no perturbation during their experiment. Unfortunately the method is ultimately unconvincing to a skeptical macrorealist, and so the conclusions drawn by SOS are unjustified.
\end{abstract}
\submitto{\NJP}
\maketitle

\section{Introduction}
As Souza, Oliveira and Sarthour (SOS) summarise \cite{SouzaOliveiraSarthour2011}, the Leggett-Garg (LG) test \cite{LeggettGarg1985} involves measuring two-time correlators $C_{k,m}:=\langle \mathcal{O}(t_k)\mathcal{O}(t_m)\rangle$ which quantify the average degree to which the observable $\mathcal{O}$ correlates with itself between time $t_k$ and time $t_m$. One may ensure the observable is dichotomic by defining $\mathcal{O}:=2|\psi_0\rangle\langle\psi_0|-\mathcal{I}$ for an initial state $|\psi_0\rangle$. Since this observable has eigenvalues $\pm1$, $C_{k,m}$ is easily obtained by measuring $\mathcal{O}$, waiting, and measuring $\mathcal{O}$ again. The outcomes of the measurement are then multiplied, and their product is averaged over many runs of the experiment. If several of these correlators are computed, one can construct e.g.
\begin{equation}
K=C_{1,2}+C_{2,3}-C_{1,3}.
\end{equation}
If the correlators are measured on many identical copies of the system, the assumptions of macrorealism and non-invasive measurabilty lead to a bound on this quantity: one can show that 
\begin{equation}
K\leq1.
\label{LGI}
\end{equation} 
LG knew that this inequality can be violated by a quantum system if the three correlators are determined in separate experiments: the reason for this being that all measurements on quantum systems are subject to a trade-off between information gain and disturbance. A fully projective measurement of a two level quantum system can provide the maximum 1 classical bit of information, but also threatens the maximally disturbing effect of updating the quantum state of the system onto an eigenstate of the measurement observable, which may be far from the original quantum state. Any future evolution of the state proceeds in general from this post-measurement eigenstate and \emph{not} from the pre-measurement state, as it would have done if no measurement were performed. This effect is at the heart of the LG test.
\section{Zero knowledge does not imply zero disturbance}

LG realized the importance of motivating the non-invasive measurabilty assumption. In contrast to a Bell inequality test, where one can arrange the measurements involved in the experiment at space-like intervals, this is impossible for the LG test. In the former case, the special theory of relativity provides a very strong reason to doubt that each measurement could have any influence on the other (due to the finite upper bound on the speed of a signal propagating between the two space-time locations concerned). In the latter case one cannot spatially separate the measurements, since they are applied to the same physical system. It is not obvious how to arrange the measurements so that a skeptical onlooker will not claim that they have disturbed the system, catastrophically corrupting the experimental data. Unless the assumption is convincingly motivated, the derivation of (\ref{LGI}) has little basis.  

%Attempts have been made to motivate NIM. Many attempts have employed the `quantum weak measurement' approach \cite{GogginAlmeidaBarbieri2011,Palacios-LMalletNguyen2010}, which can provide (an admittedly quantum mechanical) motivation for the non-invasive nature of certain measurements.  In these cases the interaction between the system of interest and the measuring apparatus is reduced. According to quantum theory, the state vector undergoes an arbitrarily small perturbation as the measurement strength is scaled down towards zero. SOS instead implement strong projective measurements on the maximally mixed state, which in this case unfortunately cannot be certified as non-invasive by a macrorealist.

It is well established that the initial state of the system of interest is not relevant to the LG test; this was pointed out in the original paper \cite{LeggettGarg1985}. SOS are thus quite justified in preparing their system in the maximally mixed state for the purpose of testing the LG inequality. They are not, however, justified in claiming that this implies that all and any measurements made on this state are non-invasive in the sense that LG intended. The interpretation of a mixed state is clear for both quantum physics and classical physics, as it expresses incomplete knowledge about the state of a system. It is true that in quantum physics there is perhaps a richer interpretation of a mixed state: it is a probability distribution on the Hilbert space. There are a multitude of convex decompositions of a mixed state into pure quantum states. For classical physics it is a probability distribution on the classical state space.  In either case the maximally mixed state represents zero information about the two-level system being investigated. In SOS's proposed quantum circuit, the state of the system remains a maximally mixed state throughout. This means that at all times there is zero information available about the state, so that the subjective description of the state will remain constant, although one suspects that the objective, physical state of affairs may be changing. In fact this is the case, if one computes the evolution of for example $|\psi_0\rangle=|0\rangle$ or $|1\rangle$ individually \emph{one finds that these states are indeed perturbed}, and moreover that they are perturbed in equal but opposite ways. We pose the question: how can our ignorance of the identity of the state (according to macrorealism it is either $|0\rangle$ or $|1\rangle$) mitigate the invasiveness of measurements?

To make this point more concrete, consider the following scenario. Alice flips a coin but is blindfolded. She ascribes the maximally mixed state to the coin, as there is an equal probability of it showing heads or tails. Now, while remaining blindfolded, Alice turns the coin over, effectively mapping heads into tails or tails into heads, depending on the physical state of the coin. This interaction with the coin is clearly potentially invasive (the coin may now behave differently to the case where no interaction had taken place), but still the state of the coin is the maximally mixed state. There is a very strong analogy between arguing in this scenario that the interaction is non-invasive, and SOS's argument that their circuit contains non-invasive measurements. This is our chief objection to SOS's approach.

The issue may be resolved in the way that LG suggest. One makes measurements \emph{which are convincingly non-invasive to a macrorealist}. Whether they are invasive according to quantum theory is irrelevant. The most convincing protocol known to us when viewed from a macrorealist viewpoint is the ideal-negative result measurement scheme espoused by LG and in particular Leggett \cite{LeggettGarg1985,Leggett1988}. These measurements directly exploit a macrorealist's belief that the system is in one state or the other at all times, and effectively measure the system without ever interacting with it. An alternative approach would be to experimentally determine an operational notion of measurement invasiveness through a series of control experiments: This approach is suggested by Wilde and Mizel~\cite{WildeMizel2011}. In both cases the quantum mechanical measurement induced disturbance is what gives rise to violation of the LG inequality. Other approaches do not require this disturbance and include for example taking on additional assumptions such as stationarity~\cite{HuelgaMarshallSantos1995}, or using a weak measurement scheme~\cite{GogginAlmeidaBarbieri2011,Palacios-LMalletNguyen2010}, which reduces the interaction strength between the system and the measuring device. These approaches may not require quantum mechanical back-action for a violation, but this is in contrast to Leggett and Garg's original proposal, and may therefore have different implications for the plausibility of macrorealist theories.

Can the experiment of SOS be adapted to include, for example, ideal negative result measurements? Possibly, but in order to be convincing the problems we describe in the following sections will have to be addressed. 

\section{Detector efficiency insufficiency}
As discussed in~\cite{KneeSimmonsGauger2012}, ideal negative result measurements can be carried out in a spin ensemble setting with a probe qubit; but this probe qubit must be initialized with high confidence. A nuclear spin at room temperature may appear to be well prepared in the pseudo-pure approximation, but is almost completely corrupted when one takes the whole ensemble into account. The decomposition of the state $\rho$ of the nuclear spin ensemble (which serves as SOS's experimental system) into a pure part $\rho_{pp}$ and a maximally mixed part $\mathcal{I}/2$ is given by  
\begin{equation}
\rho=\epsilon\rho_{pp}+(1-\epsilon)\mathcal{I}/2.
\end{equation}
SOS claim ``Since $(1-\epsilon)\mathcal{I}/2$ is not observed, the probe qubit in such a mixed state produces the same result as would be observed if the probe qubit were in a pure state and the detection efficiency of the measurement apparatus were $\epsilon$.''.  At thermal equilibrium $\epsilon=(1-\alpha)/(1+\alpha)$ with $\alpha=\textrm{exp}(-\mu_NB/kT)$. For typical values of temperature $T$ and magnetic field strength $B$, we find that $\epsilon< 10^{-7}$ (here $\mu_N$ is the magnetic moment of the probe nucleus and $k$ is Boltzmann's constant).
Assuming the quoted assertion is correct, and $\epsilon$ can be interpreted as a detector efficiency, it is rather low~\cite{GargMermin1987}.
~Experiments with low efficiency detection can only be convincing if a `fair sampling hypothesis' is justified. 

\section{The sampling is unfair}
The fair sampling hypothesis can be stated as follows: If one only measures a fraction of the systems which one has prepared, the gathered statistics faithfully represent the entire ensemble. This may be warranted, for example, in the case of an experiment with photon loss: typically there is no reason to suspect that unobserved photons would have given different results to observed photons, had they indeed been detected.  In nuclear magnetic resonance (NMR) the situation is different and the fair sampling assumption is patently false: It is generally accepted that the unobserved component of the nuclear spin ensemble behaves very differently to the measurable part. It is unobservable precisely because it generates a zero net magnetic field; the field from each spin is cancelled out by other spins in the `identity' component. If the unobserved spins behaved in the same way as the observed spins, they would become observable - giving a contradiction.  
When authoring a previous paper~\cite{SouzaMagalhaesTeles2008} (their Ref. 23) SOS and coauthors claim to have \emph{simulated} the violation of a Bell inequality with a room temperature NMR experiment - precisely because of the failure of the fair sampling hypothesis. 
In contrast, in the work under consideration here, despite the experimental system being the same, SOS do not regard their experiment as a simulation. 

One way of overcoming these difficulties is to construct the total density matrix of a highly polarised spin ensemble for analysis with the LG inequality. An experimental violation was found using this method in Ref~\cite{KneeSimmonsGauger2012}.
\section{Pitfalls of quantum circuits}
We would like to point out another reason why performing `ideal negative result measurements' on NMR systems can be tricky. Controlled-NOT (or CNOT) gates, which flip the state of an ancillary system whenever the control system is in a particular state, can be used along with postselection to implement these special measurements. In the quantum circuit paradigm, they can be built by composing several other gates, some of which may be unconditional on the control system.
\begin{figure}
\begin{center}
\includegraphics[width=12cm]{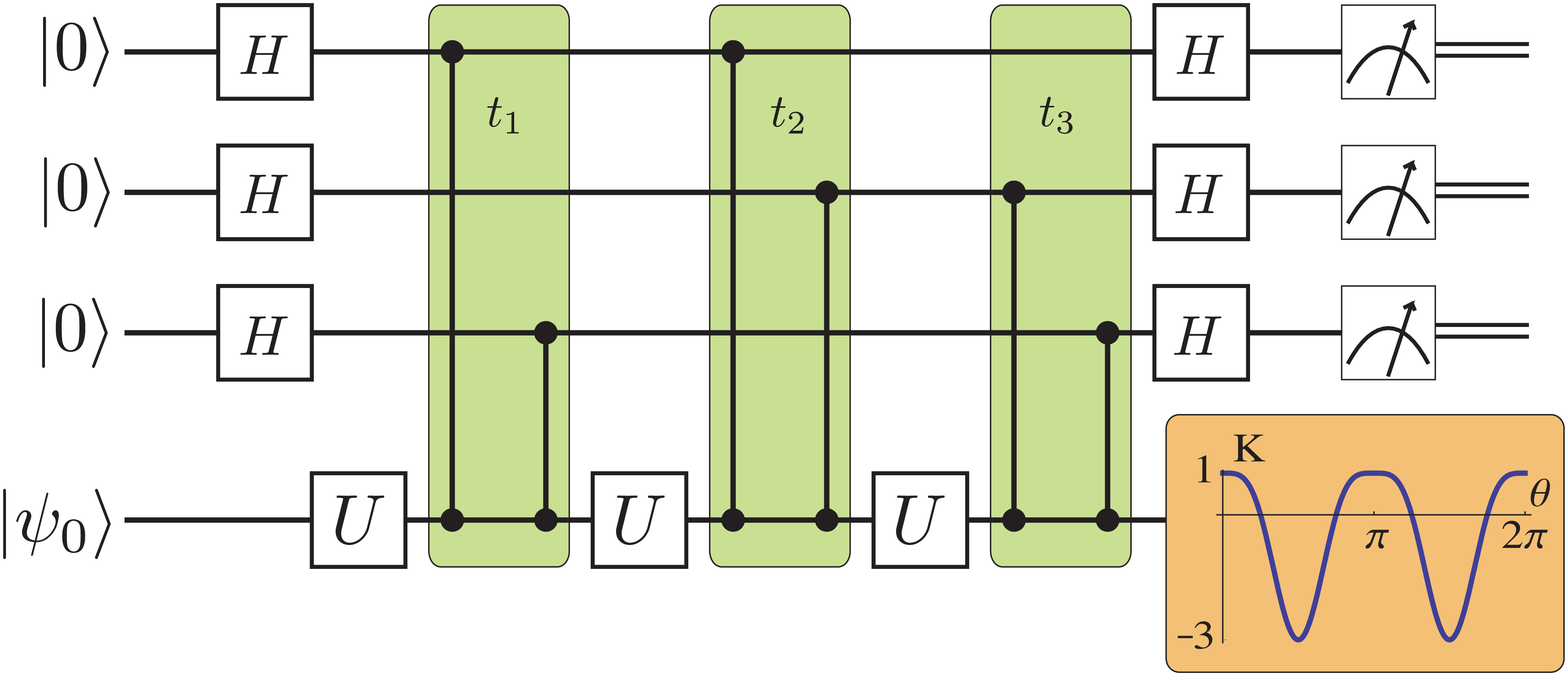}
\caption{\label{fig1} A circuit that might be suggested to test the LG inequality in a single experiment but which will fail to violate the bound on macrorealism despite the existence of coherent superpositions. Since the observable $\mathcal{O}=\sigma_z$, the controlled rotation is a controlled phase gate. At each of the instants $t_1,t_2,t_3$ two phase gates map information about $\mathcal{O}$ onto three ancillary qubits, and in the intervening times the system qubit is evolved according to $U(\theta)=\cos\theta\mathcal{I}+i\sin\theta\sigma_x$. Each of the ancillary qubits is measured to determine (after ensemble averaging) $C_{1,2}, C_{2,3},C_{1,3}$ respectively. The inset shows that the LG inequality will be obeyed for any value of $\theta$ (which is a function of $\tau:=t_3-t_2=t_2-t_1$).}
\end{center}
\end{figure}

Finally we note the final sentence of the paper: ``\ldots we would like to mention that the scattering quantum circuit presented here can be easily adapted to measure the three correlation functions simultaneously using more ancillary qubits''. We suspect that this may indeed be possible, but that a violation of the LG inequality is impossible in this case (see Figure \ref{fig1}).
In the test as LG originally outlined it, it is necessary to measure each correlator in a separate run and not `simultaneously'  (we take simultaneously here to mean `in a single run'). This is because there is no single evolution of a two-level system which is compatible with a violation of the LG inequality, whether the evolution be thought of as a classical two level system flipping from one of its states to the other, \emph{or} as a quantum system evolving under the continuous time evolution of the Schr\"odinger equation and the discontinuous back action of projective measurements.  
\section*{References}
% Create the reference section using BibTeX:

\end{document}